\documentclass[prl]{revtex4}

\begin{document}

\title{Gravitational qubits\\}

\author{Giorgio Papini}

\altaffiliation[Electronic address:]{papini@uregina.ca}
\affiliation{Department of Physics and Prairie Particle Physics Institute, University of Regina, Regina, Sask. S4S 0A2, Canada}%

\begin{abstract}
We report on the behaviour of two-level quantum systems, or qubits, in the background of
rotating and non-rotating metrics and provide a method to derive
the related spin currents and motions. The calculations are performed in the external field approximation.
\end{abstract}

\maketitle

\section{1. Introduction}

Spin effects in gravity straddle the boundary between quantum and classical physics. The
difference between quantum and classical behaviour becomes particularly transparent with
spin, which is relevant to low energy approaches to quantum gravity.

Quantum beats are an unequivocal indication that the system considered obeys the laws of quantum mechanics.
Quantum systems with a two-dimensional Hilbert space are also called qubits. This definition is borrowed
from quantum computing where two-level quantum systems play a predominant role. Qubits provide examples
of systems that are genuinely quantum mechanical and, at the same time simple, can be studied within the confines
of first quantization and are ideal in the study of relativistic gravity close to, or at the
quantum level. There are gravitational qubits in the universe. Some of them are
disussed below.

In what follows, use is made of the external field approximation \cite{PAP1,PAP0} that treats gravity as a classical
theory when it interacts with quantum particles.
This approximation can be applied successfully to all those problems
involving gravitational sources of weak to intermediate strength for which the full-fledged use of general relativity
is not required \cite{PASP,PAP10,PAP11,PAP5, LAMB,PAP6,PAP8, PAP9,PAP7}, it is encountered in the solution of relativistic wave equations and takes different forms
according to the statistics obeyed by the particles \cite{PAP0,PAP2,PAP3,PAP4,PAP5}. The approximation can also be applied
to theories in which acceleration has an upper limit \cite{CAI1,CAI2,CAI3,BRA,MASH1,MASH2,MASH3,TOLL,SCHW,PU} and that allow for the resolution
of astrophysical and cosmological singularities in quantum gravity \cite{ROV,BRU}. It is of interest to those theories of asymptotically
safe gravity that can be expressed as Einstein gravity coupled to a scalar field \cite{CA}, and
can produce results complementary to those of the method
of space-time deformation \cite{CapLamb}.

At the same time, theoretical developments by Mashhoon \cite{mashhoon1,MASH1,MASH2,mashhoon4,mashhoon2},
and by other authors \cite{hehl.ni,caipap1, papprd1,papprd2, gpap,gpap1}
in the field of spin-gravity coupling require a scheme that involves all components of the metric tensor .
Finally, recent experimental
observations of
important rotation-related classical effects \cite{Everitt, Iorio1, Iorio2}, of spin-rotation coupling for photons \cite{ashby} and neutrons \cite{demirel},
the development of a spin rotator for neutron interferometry \cite{demirel2} and the generation
of spin currents via spin-rotation coupling \cite{KYM} indicate
the degree of maturity and breadth of scope reached by the field. 

In the formalism introduced in references \cite{PAP1,PAP0}, the effect of gravity on wave functions 
is contained in a phase factor. If phase differences develop in processes involving the 
qubits studied below, measurements become possible. This is the common thread that links the various 
sections of this work. If gravitation produces qubits, then these may be observable and yield useful 
experimental results. Quantum physics and gravity may meet well before the onset of the quantum 
gravity regime usually associated with Planck's length and there still are interesting problems 
to investigate at lower scales. For instance, in addition to the important 
classical effect observed and discussed in references \cite{Everitt, Iorio1, Iorio2}, 
there also is a quantum Lense-Thirring effect, that represents the action of the Lense-Thirring metric on a 
particle wave function. By applying the procedure of \cite{PAP1,PAP0}, one finds \cite{gpap1} that 
the phase difference produced by a gravitational source of mass $M$, radius $R$ and angular velocity $\omega$ 
is $\Delta \chi_{LT}= \Omega_{LT} \Pi$, where $\Omega_{LT}=2G M \omega/(5 c^2 R)$ is the 
effective Lense-Thirring frequency of a gyroscope and $\Pi =4m \ell^{2}/\hbar$ replaces the period 
of a satellite in the classical calculation. Its observation with neutron 
interferometers of typical dimension $\ell \sim 3$cm, still seems difficult, but would 
complete nicely what we know at present about rotation in relativity.

For the sake of completeness, some essential points are being repeated.
The key player in what follows is the covariant Dirac equation
\begin{equation}\label{DiracEquation}
  [i\gamma^\mu(x){\cal D}_\mu-m]\Psi(x)=0\,,
\end{equation}
that determines the behaviour of spin-1/2 particles in the presence of a
gravitational field $g_{\mu\nu}$. In (\ref{DiracEquation}),
${\cal D}_\mu=\nabla_\mu+i\Gamma_\mu (x)$, $\nabla_\mu$ is
the covariant derivative, $\Gamma_{\mu}(x)$ the spin connection
and the matrices $\gamma^{\mu}(x)$ satisfy the relations
$\{\gamma^\mu(x), \gamma^\nu(x)\}=2g^{\mu\nu}$. Both
$\Gamma_\mu(x)$ and $\gamma^\mu(x)$ can be obtained from the usual
constant Dirac matrices $\gamma^{\hat{\alpha}}$ by using the vierbein fields $e_{\hat
\alpha}^\mu$ and the relations
\begin{equation}\label{II.2}
  \gamma^\mu(x)=e^\mu_{\hat \alpha}(x) \gamma^{\hat
  \alpha}\,,\qquad
  \Gamma_\mu(x)=-\frac{1}{4} \sigma^{{\hat \alpha}{\hat \beta}}
  e^\nu_{\hat \alpha}e_{\nu\hat{\beta};\, \mu}\,,
\end{equation}
where $\sigma^{{\hat \alpha}{\hat \beta}}=\frac{i}{2}[\gamma^{\hat
\alpha}, \gamma^{\hat \beta}]$. We use units $ \hbar = c = 1$ and the notations are as in \cite{PASP}.

Eq. (\ref{DiracEquation}) can be solved exactly \cite{PAP2,singh} to first order in
the metric deviation
$\gamma_{\mu\nu}(x)=g_{\mu\nu}-\eta_{\mu\nu}$, where the Minkowski
metric $\eta_{\mu\nu}$ has signature -2. This is achieved by first
transforming (\ref{DiracEquation}) into the equation
\begin{equation}\label{DiracEqTrasf}
  [i{\tilde \gamma}^\nu (x) \nabla_{\nu}-m]{\tilde \Psi}(x)=0\,,
\end{equation}
where
\begin{equation}\label{PsiTilde}
{\tilde \Psi}(x)=S^{-1}\Psi(x)\,,\qquad S(x)=e^{-i\Phi_s(x)}\,,
\qquad \Phi_s(x)={\cal P}\int_P^x dz^\lambda \Gamma_\lambda (z)\,,
\qquad {\tilde \gamma}^{\mu}(x)=S^{-1}\gamma^{\mu}(x) S\,.
\end{equation}
By multiplying (\ref{DiracEqTrasf}) on the left by $(-i{\tilde
\gamma}^\nu (x)\nabla_{\nu}-m)$, we obtain the equation
\begin{equation}\label{KGequation}
  (g^{\mu\nu}\nabla_\mu\nabla_\nu+m^2){\tilde \Psi}(x)=0\,,
\end{equation}
whose solution
\begin{equation}\label{ExactSolution}
  {\tilde \Psi}(x)=e^{-i\hat{\Phi}_G(x)}\Psi_0(x)\,,
\end{equation}
is exact to first order. The operator $\hat{\Phi}_G(x)$ is defined
as
\begin{equation}\label{PhiG}
  \hat{\Phi}_{G}=-\frac{1}{4}\int_P^xdz^\lambda\left[\gamma_{\alpha\lambda,
  \beta}(z)-\gamma_{\beta\lambda, \alpha}(z)\right]\hat{L}^{\alpha\beta}(z)+
  \frac{1}{2}\int_P^x dz^\lambda\gamma_{\alpha\lambda}\hat{k}^\alpha\,,
\end{equation}
\[
 [\hat{L}^{\alpha\beta}(z), \Psi_0(x)]=\left((x^\alpha-z^\alpha)\hat{k}^\beta-
 (x^\beta-z^\beta)\hat{k}^\alpha\right)\Psi_0(x)\,, \qquad
 [\hat{k}^\alpha, \Psi_0(x)]=i\partial^\alpha\Psi_0\,,
 \]
and $\Psi_0(x)$ satisfies the usual free Dirac equation
\begin{equation}\label{DE}
\left(i\gamma^{\hat{\mu}}\partial_{\mu}-m\right)\Psi_{0}(x)=0\,.
\end{equation}
In (\ref{PsiTilde}) and (\ref{PhiG}), the path integrals are taken along
the classical world line of the particle starting from an arbitrary reference point
$P$. Only the path to $\mathcal{O}(\gamma_{\mu\nu})$ needs to be known in the integrations indicated
because (\ref{PsiTilde}) already is a first order solution. The positive energy solutions of (\ref{DE}) are given by
\begin{equation}\label{psi0}
\Psi_{0}(x)=u({\bf k})e^{-ik_\alpha x^\alpha}=N
  \left(\begin{array}{c}
                \phi \\
                 \frac{{\bf \sigma}\cdot{\bf k}}{E+m}\, \phi \end{array}\right)
                 \,e^{-ik_\alpha x^\alpha}\,,
\end{equation}
where $N=\sqrt{\frac{E+m}{2E}}$, $u^{+} u=1$, $\bar{u}=u^{+}\gamma^{0}$, $u^{+}_{1}u_{2}=u_{2}^{+}u_{1}=0$ and ${\bf \sigma}=(\sigma^1, \sigma^2, \sigma^3)$ represents the
Pauli matrices.. In addition $\phi$ can take the forms $ \phi_{1}$  and $\phi_{2}$ where
$\phi_{1}=\left(\begin{array}{c}
1\\ 0 \end{array}\right)$, and $\phi_{2}=\left(\begin{array}{c}
0\\ 1 \end{array}\right)$.

$\hat{L}_{\alpha\beta}$ and $\hat{k}^\alpha$ are the angular and
linear momentum operators of the particle. It follows from
(\ref{ExactSolution}) and (\ref{PsiTilde}) that the solution of
(\ref{DiracEquation}) can be written in the form \cite{PAP5}
\begin{equation}\label{PsiSolution}
  \Psi(x)=e^{-i\Phi_s}\left(-i{\tilde \gamma}^\mu(x)\nabla_\mu
  -m\right)e^{-i\Phi_G}\, \Psi_0(x)\equiv \hat{T}\Psi_{0}\,,
\end{equation}
and also as
\begin{equation}\label{PsiSolution2}
  \Psi(x)=-\frac{1}{2m}\left(-i\gamma^\mu(x){\cal
  D}_\mu-m\right)e^{-i\Phi_T}\Psi_0(x)\equiv \hat{T}\Psi_{0}\,,
\end{equation}
where $\Phi_T=\Phi_s+\Phi_G$ is of first order in
$\gamma_{\alpha\beta}(x)$. The condition that both sides
of the equation agree when the gravitational field vanishes
accounts for the presence of the factor $ -1/2m $ on the r.h.s. of
(\ref{PsiSolution2}).

On multiplying (\ref{DiracEquation}) on the left
by $(-i\gamma^\nu(x){\cal D}_\nu-m)$ and using the relations
\begin{equation}\label{relation1}
  \nabla_\mu\Gamma_\nu(x)-\nabla_\nu\Gamma_\mu(x)+i[\Gamma_\mu(x),
  \Gamma_\nu(x)]=-\frac{1}{4}\sigma^{\alpha\beta}(x)R_{\alpha\beta\mu\nu}\,,
\end{equation}
and
\begin{equation}\label{relation2}
  [{\cal D}_\mu, {\cal D}_\nu]=-\frac{i}{4}\,
  \sigma^{\alpha\beta}(x)R_{\alpha\beta\mu\nu}\,,
\end{equation}
we obtain the equation
\begin{equation}\label{KGEq+R}
  \left(g^{\mu\nu}{\cal D}_\mu{\cal
  D}_\nu-\frac{R}{4}+m^2\right)\Psi(x)=0\,.
\end{equation}
In (\ref{relation2}) and (\ref{KGEq+R}), $R_{\alpha\beta\lambda\mu}(z)=-\frac{1}{2}\left(\gamma_{\alpha\lambda,\beta\mu}
+\gamma_{\beta\mu,\alpha\lambda}-\gamma_{\alpha\mu,\beta\lambda}-\gamma_{\beta\lambda,\alpha\mu}\right)$
 is the linearized Riemann tensor, $R$ the corresponding Ricci scalar, and
$\sigma^{\alpha\beta}(x)=(i/2)[\gamma^\alpha(x),
\gamma^\beta(x)]$.

By using Eq. (\ref{PsiTilde}), we also find
\begin{equation}\label{Solution}
  (-i\gamma^\nu(x){\cal D}_\nu-m)\,S\, (i{\tilde
  \gamma}^\mu\nabla_\mu-m){\tilde \Psi}(x)=
  S\, (g^{\mu\nu}\nabla_\mu\nabla_\nu+m^2){\tilde \Psi}(x)=0\,.
\end{equation}
Eq. (\ref{KGEq+R}) implies that the gyro-gravitational ratio of a
massive Dirac particle is one, as found in \cite{oliveira,audretsch,kannenberg}.

The transformations of coordinates $
x_{\mu}\rightarrow x_{\mu}+\xi_{\mu}$, with $ \xi_{\mu}(x) $
small of first order, lead to the "gauge"
transformations $ \gamma_{\mu\nu}\rightarrow
\gamma_{\mu\nu}-\xi_{\mu,\nu}-\xi_{\nu,\mu}$. It is therefore
necessary to show that  $\Phi_{T}$ in (\ref{PsiSolution2}) is
gauge invariant. In fact, on applying Stokes theorem to a closed
spacetime path $C$ and using (\ref{relation1}), we find that
$\Phi_T$ changes by
\begin{equation}\label{phase}
  \Delta \Phi_T=\frac{1}{4}\int_\Sigma d\tau^{\mu\nu}J^{\alpha\beta}
  R_{\mu\nu\alpha\beta}\,,
\end{equation}
where $\Sigma$ is a surface bound by $C$ and $J^{\alpha\beta}$ is
the total angular momentum of the particle. Eq. (\ref{phase}) shows that (\ref{PsiSolution})
and (\ref{PsiSolution2}) are gauge invariant and confirms that, to
first order in the gravitational field, the gyro-gravitational
ratio of a Dirac particle is one. Use of
(\ref{PsiSolution}), or (\ref{PsiSolution2}) assures the correct treatment of both spin and
angular momentum.

The plan of this work is as follows. In Section 2 we
discuss qubits represented by particles in accelerators. In Section 3
we derive the gravitational deflection of particles propagating in a
gravitational background represented by the Lense-Thirring metric and
obtain the contribution due to the rotation of the source.
The neutrino helicity transitions are derived in Section 4 and some
astrophysical consequences are discussed in Section 5. Spin currents and
spin motion are presented in Section 6 and are followed by a summary.

\section{2. Spin-rotation coupling in accelerators}

The spin-rotation effect
described by Mashhoon is conceptually important,
it extends our knowledge of rotational inertia to
the quantum level and violates the principle of equivalence \cite{mashhoon1,mashhoon2} that is
well-tested at the classical level.

It has, of course been argued that the principle of equivalence does not hold true in the
quantum world. This is the case for phase shifts in particle
interferometers \cite{lammer,singh} and wave functions depend on the masses of
the particles involved \cite{greenberg}. In addition, the equivalence principle
does not apply in the context of the causal interpretation of quantum mechanics
as shown by Holland \cite{holl}. Several models predicting quantum violations of the
equivalence principle have also been discussed in the literature \cite{morgan,peres},
also in connection with neutrino oscillations
\cite{gasperini,halprin,butler,bozza,adunas}. The Mashhoon term, in particular, yields
different potentials for different particles and for different spin states
and can not, therefore, be regarded as universal. It plays, nonetheless, an
essential role in precise measurements of the $g-2$ factor of the muon.

The experiment \cite{bailey,farley} involves muons in a storage ring. Muons on
equilibrium orbits within a small fraction of the maximum momentum are almost completely
polarized with spin vectors pointing in the direction of motion. As the muons decay,
those electrons projected forward in the muon rest frame are detected around the ring.
Their modulated angular distribution reflects the precession of the muon spin along the
cyclotron orbits.

Our calculations use the covariant Dirac equation and are performed in the rotating frame of the muon and do not therefore require
a relativistic treatment of inertial spin effects \cite{ryder}  . Then the vierbein
formalism yields $\Gamma_i=0$ and
\begin{equation}\label{1}
  \Gamma_0=-\frac{1}{2}\,
  a_i\sigma^{0i}-\frac{1}{2}\,\omega_i\sigma^i\,,
\end{equation}
where $a_i$ and $\omega_i$ are the three-acceleration and three-rotation of the observer
and
 \[
 \sigma^{0i}\equiv\frac{i}{2}\, [\gamma^0, \gamma^i]=i
 \left(\matrix{ \sigma^i & 0 \cr
                0 & -\sigma^i \cr }\right)\,
 \]
in the chiral representation of the usual Dirac matrices. The second term in (\ref{1})
represents the Mashhoon effect. The first term drops out. The remaining contributions to
the Dirac Hamiltonian, to first order in $a_i$ and $\omega_i$, are
\cite{hehl.ni,singh}
\begin{eqnarray}\label{hamiltonian-Ni}
  H &\approx & {\vec \alpha}\cdot {\vec p}+m\beta+\frac{1}{2}
  [({\vec a}\cdot {\vec x})({\vec p}\cdot {\vec \alpha})+
  ({\vec p}\cdot {\vec \alpha})({\vec a}\cdot {\vec x})] \\
  & & -{\vec \omega}\cdot \left({\vec L}+\frac{{\vec \sigma}}{2}\right)\,.
  \nonumber
\end{eqnarray}
All quantities in $H$ are time-independent and are referred
to a left-handed set of three axes rotating  about the $x_2$-axis in the clockwise direction
of motion of the muons. The muon momentum is directed along the $x_3$-axis which is tangent
to the muon orbits. The magnetic field is $B_2=-B$. Only the Mashhoon term then couples
the helicity states of the muon. The remaining terms contribute to the overall energy $E$
of the states, and we indicate by $H_0$ the corresponding part of the Hamiltonian.

Before decay the muon states can be represented as
\begin{equation}\label{2}
  |\psi(t)>=a(t)|\psi_+>+b(t)|\psi_->\,,
\end{equation}
where $|\psi_+>$ and $|\psi_->$ are the right and left helicity states of the Hamiltonian
$H_0$ and satisfy the equation
 \[
 H_0|\psi_{+,-}>=E|\psi_{+,-}>\,.
 \]
The total effective Hamiltonian is $H_{eff}=H_0+H'$, where
\begin{equation}\label{3}
  H'=-\frac{1}{2}\,\omega_2\sigma^2+\mu B\sigma^2\,.
\end{equation}
$\displaystyle{\mu=\left(1+\frac{g-2}{2}\right)\mu_0}$ represents the total magnetic
moment of the muon and $\mu_0$ is the Bohr magneton. We will neglect the presence of
electric fields that also affect the muon spin. Their effects
can be controlled in suitable ways \cite{farley}.

The coefficients $a(t)$ and $b(t)$ in (\ref{2}) evolve in time according to
\begin{equation}\label{4}
  i\frac{\partial}{\partial t} \left(\matrix{ a(t) \cr
                b(t) \cr }\right)=M \left(\matrix{ a(t) \cr
                b(t) \cr }\right)\,,
\end{equation}
where $M$ is the matrix
\begin{equation}\label{5}
  M= \left[\matrix{ E-i\displaystyle{\frac{\Gamma}{2}} &
            \displaystyle{i\left(\frac{\omega_2}{2}-\mu B\right)}\cr
                \displaystyle{-i\left(\frac{\omega_2}{2}-\mu B\right)} &
                E-i\displaystyle{\frac{\Gamma}{2}} \cr }\right]
\end{equation}
and $\Gamma$ represents the width of the muon and is not particularly relevant to what follows.
Equations (\ref{4}) and (\ref{5}) describe
a two-dimensional qubit. The non-diagonal form of $M$ (when $B=0$)
implies that rotation does not couple universally to matter.

 $M$ has eigenvalues
\begin{eqnarray}
 h_1 &=& E-i\frac{\Gamma}{2}+\frac{\omega_2}{2}-\mu B \,, \nonumber \\
 h_2 &=& E-i\frac{\Gamma}{2}-\frac{\omega_2}{2}+\mu B \,, \nonumber
\end{eqnarray}
and eigenstates
\begin{eqnarray}
 |\psi_1> &=&
 \frac{1}{\sqrt{2}}\,\left[i|\psi_+>+|\psi_->\right]\,, \nonumber
 \\
 |\psi_2> &=& \frac{1}{\sqrt{2}}\,\left[-i|\psi_+>+|\psi_->\right]
 \,. \nonumber
\end{eqnarray}
The muon states that satisfy (\ref{4}), and the condition $|\psi (0)>=|\psi_->$ at $t=0$,
are
\begin{eqnarray} \label{6a}
 |\psi(t)> &=& \frac{e^{-\Gamma t/2}}{2} e^{-iEt}
 \left\{
 i\left[e^{-i{\tilde \omega} t}
 -e^{i{\tilde \omega} t}\right]|\psi_+> \right.
 \\
 & & \left. + \left[e^{-i{\tilde \omega} t}
 +e^{i{\tilde \omega} t}\right]|\psi_-> \right\}
 \,, \nonumber
\end{eqnarray}
where
 \[
 {\tilde \omega}\equiv \frac{\omega_2}{2}-\mu B\,.
 \]
The spin-flip probability is therefore
\begin{eqnarray}\label{7}
  P_{\psi_-\to \psi_+}&=&|<\psi_+|\psi(t)>|^2 \\
     &=& \frac{e^{-\Gamma
  t}}{2}[1-\cos(2\mu B-\omega_2) t]\,. \nonumber
\end{eqnarray}
The $\Gamma$-term in (\ref{7}) accounts for the observed exponential decrease in electron
counts due to the loss of muons by radioactive decay \cite{farley}. The term in square
brackets represents the well known
phenomenon of quantum beats which one should expect because muons are quantum systems.
It also represents the characteristic behaviour of a two-dimensional qubit.

The spin-rotation contribution to $P_{\psi_-\to \psi_+}$ is represented by $\omega_2$
which is the cyclotron angular velocity $\displaystyle{\frac{eB}{m}}$ \cite{farley}. The
spin-flip angular frequency is then
 \begin{eqnarray}\label{omegafin}
 \Omega&=&2\mu B-\omega_2 \\
 &=&\left(1+\frac{g-2}{2}\right)\frac{eB}{m}-
 \frac{eB}{m} \nonumber \\
 &=& \frac{g-2}{2}\frac{eB}{m}\equiv \frac{1}{\Delta}\,, \nonumber
 \end{eqnarray}
which is precisely the observed modulation frequency of the electron counts
\cite{picasso} (see also Fig. 19 of Ref. \cite{farley}) and yields the value $\Delta$
of the energy level splitting. This result is independent of
the value of the anomalous magnetic moment of the particle. It is therefore the spin-rotation
coupling that gives evidence to the $g-2$ term in $\Omega$ by exactly cancelling, in $2\mu B$, the
much larger contribution $\mu_0$ that one would get if the fermion had no anomalous magnetic
moment. The cancellation is made possible by the non-diagonal form of $M$ and is
therefore a direct consequence of the violation of the equivalence principle.

It is perhaps surprising that spin-rotation coupling as such has almost gone unnoticed for such
a long time. It is however significant that its effect is observed in an experiment that
has already provided crucial tests of quantum electrodynamics  and a test of Einstein's
time-dilation formula to better than a 0.1 percent accuracy.

Applications of these ideas to compound spin systems like heavy ions in accelerators
can be found in \cite{GSI1,GSI2,GSI3}.

\section{3. Geometrical Optics of Spin-1/2 Particles}

In this Section we study the propagation of a spin-1/2 particle
in the Lense-Thirring metric \cite{Lense}
represented, in its post-Newtonian form, by
\begin{equation}\label{LTmetric}
 \gamma_{00}=2\varphi\,, \quad \gamma_{ij}=2\varphi\delta_{ij}\,, \quad
\gamma_{0i}=h_i=\frac{2}{r^3}(\textbf{J}\wedge\textbf{r})_i\,,
 \end{equation}
where
\begin{equation}\label{LTmetric1}
  \varphi=-\frac{GM}{r}\,, \quad
  \textbf{h}=\frac{4GMR^{2}\omega}{5r^3}(y,-x,0)\,,
\end{equation}
and  \textit{M}, \textit{R}, $ \mathbf{\omega}=(0,0,\omega)$ and $
\mathbf{J}$ are mass, radius, angular velocity and angular
momentum of the source. The vierbein field to
$\mathcal{O}(\gamma_{\mu\nu})$ is
\begin{equation}\label{3.5}
 e^0_{\hat{i}}=0\,{,}\quad
 e^0_{\hat{0}}=1-\varphi\,{,}\quad
 e^i_{\hat{0}}=h_i\,{,}\quad
 e^l_{\hat{k}}=\left(1+\varphi\right)\delta^l_k\,.
 \end{equation}
It can further isolate the gravitational
contribution in (\ref{3.5}) by writing
 $e^{\mu}_{\hat{\alpha}}\simeq\delta^{\mu}_{\hat{\alpha}}+h^{\mu}_{\hat{\alpha}}$.
 The components of the spin connection can be calculated using (\ref{II.2}) and
(\ref{3.5}) and are
\begin{eqnarray}\label{3.6}
  \Gamma_0 &=& -\frac{1}{2}\varphi,_j\sigma^{{\hat 0}{\hat j}}
 -\frac{1}{8}(h_{i,j}-h_{j,i})\sigma^{{\hat i}{\hat j}} \\
  \Gamma_i &=& -\frac{1}{8}(h_{i,j}-h_{j,i})\sigma^{{\hat 0}{\hat j}}
            -\frac{1}{2}\varphi,_j\sigma^{{\hat i}{\hat j}}\nonumber\,,
 \end{eqnarray}
and have the explicit form
\begin{eqnarray}\label{spinexpl}
  \Gamma_0&=&-\frac{GM}{2r^3}\left(x\sigma^{\hat{0}\hat{1}}+y\sigma^{\hat{0}\hat{2}}+z\sigma^{\hat{0}\hat{3}}\right)+\frac{GMR^2\omega}{5r^5}\left[(r^2-3z^2)\sigma^{\hat{1}\hat{2}}+3yz\sigma^{\hat{1}\hat{3}}-3xz\sigma^{\hat{2}\hat{3}}\right]\\
  \Gamma_1&=&\frac{3GMR^2\omega}{5r^5}\left[2xy\sigma^{\hat{0}\hat{1}}+(y^2-x^2)\sigma^{\hat{0}\hat{2}}+yz\sigma^{\hat{0}\hat{3}}\right]+\frac{GM}{2r^3}\left(y\sigma^{\hat{1}\hat{2}}+z\sigma^{\hat{1}\hat{3}}\right)\nonumber\\
  \Gamma_2&=&\frac{3GMR^2\omega}{5r^5}\left[(y^2-x^2)\sigma^{\hat{0}\hat{1}}-2xy\sigma^{\hat{0}\hat{2}}-xz\sigma^{\hat{0}\hat{3}}\right]+\frac{GM}{2r^3}\left(-x\sigma^{\hat{1}\hat{2}}+z\sigma^{\hat{2}\hat{3}}\right)\nonumber\\
  \Gamma_3&=&\frac{3GMR^2\omega}{5r^5}\left(yz\sigma^{\hat{0}\hat{1}}-xz\sigma^{\hat{0}\hat{2}}\right)+\frac{GM}{2r^3}\left(x\sigma^{\hat{1}\hat{3}}+y\sigma^{\hat{2}\hat{3}}\right)\,.\nonumber
\end{eqnarray}

In what follows, use is made of the Dirac
representation of the $\gamma^{\hat{\mu}}$, of the first derivative of $\Phi_{G}$ with respect
to $x^\mu$
\begin{equation}\label{BPhiGDer}
  \Phi_{G, \mu}=-\frac{1}{2}\int_P^x dz^\lambda (\gamma_{\mu\lambda,
  \beta}-\gamma_{\beta\lambda,
  \mu})k^\beta+\frac{1}{2}\gamma_{\alpha\mu}k^\alpha\,,
\end{equation}
and of the second derivative
\begin{equation}\label{PhiGDDerv}
  \Phi_{G, \mu\nu}=k_\alpha \Gamma^\alpha_{\mu\nu}\,,
\end{equation}
where $\Gamma^\alpha_{\mu\nu}$ are the Christoffell symbols of the
second type.\\
For the Lense-Thirring metric and to
$ \mathcal{O}(\gamma_{\mu\nu})$, these are
\begin{eqnarray}\label{Christ}
  \Gamma^0_{00}&=&0\,,\quad\Gamma^0_{0i}=\varphi_{,i}\,,\quad\Gamma^0_{ij}=\frac{1}{2}(h_{i,j}+h_{j,i})\,,\\
  \Gamma^i_{00}&=&\varphi_{,i}\,,\quad\Gamma^i_{0j}=\frac{1}{2}(h_{j,i}-h_{i,j})\,,\quad\Gamma^i_{jk}=\delta^j_k\varphi_{,i}-\delta^i_j\varphi_{,k}-\delta^i_k\varphi_{,j}\,.\nonumber
\end{eqnarray}

 In the geometrical optics
approximation $|\partial_i\gamma_{\mu\nu}|\ll|k\gamma_{\mu\nu} |$, where $k$ is
the momentum of the particle, the geometrical
phase $\Phi_G$ is sufficient to reproduce the classical angle of
deflection, as it should, but also some
effects due to the angular velocity of rotation of the source.

The deflection angle $\delta$ is defined by
\begin{equation}\label{tgangle}
  \tan \delta = \frac{\sqrt{-g_{ij}p_\perp^i
  p_\perp^j}}{p_\parallel}\simeq \frac{|{\bf p}_\perp|}{k_\parallel}\,,
\end{equation}
where $k_\parallel = p_\parallel$ is the unperturbed momentum and
$|{\bf p}_\perp|=\sqrt{-\eta_{ij}p^i_\perp p^j_\perp}$, for
$p_\perp^i \sim {\cal O}(\gamma_{\mu\nu})$.

It follows from (\ref{PhiG}) and (\ref{PsiSolution}) that, once
$\Psi_0(x)$ is chosen to be a plane wave solution of the flat
spacetime Dirac equation, the geometrical phase of a particle of
four-momentum $ k^{\mu} $ is given by
\begin{equation}\label{A1}
  \upsilon(x)= -k_\alpha x^\alpha -\Phi_G(x)\,,
\end{equation}
where $\hat{\Phi}_{G}\Psi_{0}=\Phi_{G}\Psi_{0}$ and
\begin{equation}\label{PhiGtilde}
  \Phi_G(x)=-\frac{1}{4}\int_P^xdz^\lambda\left[\gamma_{\alpha\lambda,
  \beta}(z)-\gamma_{\beta\lambda,
  \alpha}(z)\right]((x^{\alpha}-z^{\alpha})k^{\beta}-(x^{\beta}-z^{\beta})k^{\alpha})+
  \frac{1}{2}\int_P^x dz^\lambda\gamma_{\alpha\lambda}k^\alpha\,.
\end{equation}

The components of ${\bf p}_\perp$ can be determined from the
equation

\begin{eqnarray}\label{A2}
  p_i = \frac{\partial\upsilon}{\partial x^i}&=&
  -k_i - \Phi_{G, i}= \\
   & = &  -k_i-\frac{1}{2}\,\gamma_{\alpha i}(x)k^\alpha
   +\frac{1}{2}\, \int_P^x
   dz^\lambda(\gamma_{i\lambda,\beta}(z)-\gamma_{\beta\lambda,i}(z))k^\beta\,.
   \nonumber
\end{eqnarray}

We consider the two cases of propagation along the $z$-axis,
which is parallel to the angular momentum of the source, and along
the $x$-axis, orthogonal to it. In both instances, the fermions
are assumed to be ultrarelativistic, i.e. $dz^0\simeq
dz(1+m^2/2E^2) , E\simeq k(1+m^2/2E^2)$.

When motion is along the $z$-direction $\phi_{1}=\left(\begin{array}{c} 1\\
0 \end{array}\right)$ and $\phi_{2}=\left(\begin{array}{c} 0\\
1 \end{array}\right)$ (i.e. $\phi_{1,2}$ are eigenstates of
$\sigma^3$).

We consider fermions starting from
$z=-\infty$ with impact parameter $b\geq R$ and propagating along
$x=b$, $y=0$. We find
\begin{eqnarray}\label{p1z}
  p_1 &=& -\frac{1}{2}\left[\int_{-\infty}^zdz^0 \gamma_{00,1}k^0 + \int_{-\infty}^zdz^3 \gamma_{33,1} k^3\right]\\ &=&
  -2k\left(1+\frac{m^2}{2E^2}\right)\int_{-\infty}^z\varphi_{,1}dz
   \nonumber\,,\\
  p_2 &=&-\frac{1}{2}\gamma_{02}k^0
  +\frac{1}{2}\int_{-\infty}^zdz^0\gamma_{20,3}k^3=0\,\nonumber\
\end{eqnarray}
and
\begin{eqnarray}\label{pperp}
  (p_{\bot})^1&=&g^{1\mu}p_{\mu}\simeq-p_1=-\frac{2GMk}{b}\left(1+\frac{m^2}{2E^2}\right)\left(1+\frac{z}{r}\right)\,,\\
  (p_{\bot})^2&=&g^{2\mu}p_{\mu}\simeq h_2E=-\frac{4GMR^2\omega
  bk}{5r^3}\left(1+\frac{m^2}{2E^2}\right)\nonumber\,.
\end{eqnarray}

We finally obtain
\begin{equation}\label{delta0z}
  \delta =
  \frac{2GM}{b}\left(1+\frac{m^2}{2E^2}\right)\sqrt{\left(1+\frac{z}{r}\right)^2+
  \left(\frac{2R^2b^2\omega}{5r^3}\right)^2}\,,
\end{equation}
which is the deflection predicted by general relativity for
photons, with corrections due to the fermion mass and to $\omega$.
In the limit $z\to \infty$ Eq.
(\ref{delta0z}) reduces to
\begin{equation}\label{delta0z1}
 \delta = \frac{4GM}{b}\left(1+\frac{m^2}{2E^2}\right)\,.
\end{equation}

When the fermions propagate along $x$, the deflection angle is
\begin{equation}\label{delta0x}
  \delta = \frac{2GM}{b}\left(1-\frac{2R^2\omega}{5b}\right)
  \left(1+\frac{m^2}{2E^2}\right)\left(1+\frac{x}{r}\right)\,.
\end{equation}
The first term is just that predicted by general relativity.

Contrary to the case of propagation along $z$, the contribution of $\omega$
does not vanish in the limit $x\to \infty$.
In fact, in this limit we get
\begin{equation}\label{delta0x1}
  \delta = \frac{4GM}{b}\left(1-\frac{2R^2\omega}{5b}\right)
  \left(1+\frac{m^2}{2E^2}\right)\,.
\end{equation}

\section{4. Neutrino helicity transitions}

In what follows, it is convenient to write the left and right
neutrino wave functions in the form
\begin{equation}\label{Psi0}
 \Psi_0 (x) = \nu_{0L,R}e^{-ik_\alpha x^\alpha}=\sqrt{\frac{E+m}{2E}}
  \left(\begin{array}{c}
                \nu_{L, R} \\
                 \frac{{\bf \sigma}\cdot {\bf k}}{E+m}\, \nu_{L, R} \end{array}\right)
                 \,e^{-ik_\alpha x^\alpha}\,,
\end{equation}
where $\bbox \sigma=(\sigma^1, \sigma^2, \sigma^3)$ represents the
Pauli matrices. $\nu_{L,R}$ are eigenvectors of $({\bf
\sigma\cdot k})$ corresponding to negative and positive
helicity and ${\bar \nu}_{0\, L, R}(k)\equiv \nu_{0\, L,
R}^\dagger (k)\gamma^{\hat 0}$, ${\nu}_{0\, L, R
}^\dagger(k)\nu_{0\, L, R}(k)=1$. This notation already takes into
account the fact that if $ \nu_{\pm}$ are the helicity states,
then we have $ \phi_{2}\simeq \nu_{-}, \, \phi_{1}\simeq \nu_{+}$
for relativistic neutrinos. The propagation is {\it in vacuo}.

In general, the spin precesses during the motion of the neutrino.
This can be expected because of the presence of $ \Phi_{s}$
in $ \Phi_{T}$.

We now study the helicity flip of one flavour neutrinos as they
propagate in the gravitational field produced by a rotating mass.
The neutrino state vector can be written as
\begin{equation}\label{neutrinoRL}
 |\psi(\lambda) \rangle = \alpha(\lambda) |\nu_R\rangle
 +\beta(\lambda) |\nu_L\rangle\,,
\end{equation}
where $|\alpha|^2+|\beta|^2=1$ and $\lambda$ is an affine
parameter along the world-line. In order to determine $\alpha $
and $\beta $, we can write Eq. (\ref{PsiSolution}) as

\begin{equation}\label{PsiSolution3}
  |\psi(\lambda)\rangle = {\hat T}(\lambda) |\psi_0(\lambda)\rangle\,,
\end{equation}
where
\begin{equation}\label{Texpression}
  {\hat T}=-\frac{1}{2m}\left(-i\gamma^\mu(x){\cal
  D}_\mu-m\right)e^{-i\Phi_T}\,,
\end{equation}
and $|\psi_0(\lambda)\rangle $ is a plane wave solution of (\ref{psi0}). The latter can be written as
\begin{equation}\label{psizero}
|\psi_{0}(\lambda)\rangle = e^{-ik\cdot x}
\left[\alpha(0)|\nu_{R}\rangle + \beta(0)|\nu_{L}\rangle
\right]\,.
\end{equation}
$ |\psi(\lambda)\rangle $ should also be
normalized. However, this is unnecessary, because it is shown below that $ \alpha(\lambda) $ is
already of $ \mathcal{O}(\gamma_{\mu\nu})$ and can only produce
higher order terms. From (\ref{neutrinoRL}), (\ref{PsiSolution3}) and
(\ref{psizero}) we obtain
\begin{equation}\label{alpha}
\alpha(\lambda)=\langle
\nu_{R}|\psi(\lambda)\rangle=\alpha(0)\langle\nu_{R}|\hat{T}|\nu_{R}\rangle
+ \beta(0)\langle \nu_{R}|\hat{T}|\nu_{L}\rangle\,.
\end{equation}
An equation for $ \beta $ can be derived in an entirely similar
way.

If we consider neutrinos which are created in the left-handed
state, then $|\alpha(0)|^2=0,  |\beta(0)|^{2}=1 $, and we obtain
\begin{equation}\label{PLR}
  P_{L\rightarrow R}=|\alpha(\lambda)|^2=\left|\langle \nu_{R}|\hat{T}|\nu_{L}\rangle\right|^{2}
  =\left|\int_{\lambda_0}^\lambda\langle \nu_R| {\dot x}^\mu \partial_\mu {\hat
  T}|\nu_L\rangle d\lambda\right|^2\\,
\end{equation}
where $\dot{x}^{\mu}=k^{\mu}/m$. As remarked in \cite{CAR}, $
\dot{x}^{\mu}$ need not be a null vector if we assume that the
neutrino moves along an "average" trajectory. We also find, to
lowest order,
\begin{eqnarray}\label{deT}
  \partial_{\mu}\hat{T}&=&\frac{1}{2m}\left(-i2m\Phi_{G,\mu}-i(\gamma^{\hat{\alpha}}k_{\alpha}+m)\Phi_{s,\mu}
  +\gamma^{\hat{\alpha}}(h^{\beta}_{\hat{\alpha},\mu}k_{\beta}+\Phi_{G,\alpha\mu})\right)\\
  \Phi_{s,\lambda}&=&\Gamma_{\lambda}\,,\quad
  \Phi_{G,\alpha\mu}=k_{\beta}\Gamma^{\beta}_{\alpha\mu}\,,\quad
  \nu^{\dagger}_0(\gamma^{\hat{\alpha}}k_{\alpha}+m)=2E\nu^{\dagger}_{0}\gamma^{\hat{0}}\nonumber\,,
\end{eqnarray}
where $\Gamma^{\beta}_{\alpha\mu}$ are the usual Christoffel
symbols, and
\begin{equation}\label{RTL}
 \langle \nu_R |{\dot x}^\mu \partial_\mu {\hat T}|\nu_L \rangle
 =\frac{E}{m}\left[
 -i \, \frac{k^\lambda}{m} {\bar \nu}_R\Gamma_\lambda \nu_L+
 \frac{k^\lambda k_\mu}{2mE}\, (h^\mu_{{\hat \alpha},\,
 \lambda}+\Gamma^\mu_{\alpha\lambda})\nu_R^\dagger \gamma^{\hat
 \alpha}\nu_L\right].
\end{equation}
In order to solve the evolution equations for $\alpha$ and $\beta$
and complete the equations describing this two-dimensional qubit,
one also needs the terms $\left\langle\nu_{L}|\dot{x}^{\mu}\partial_{\mu}\hat{T}|\nu_{L}\right\rangle$
and $\left\langle\nu_{R}|\dot{x}^{\mu}\partial_{\mu}\hat{T}|\nu_{R}\right\rangle$ of
the usual qubit matrix $M$.
In what follows, we compute the probability amplitude (\ref{RTL})
for neutrinos propagating along the $z$ and the $x$ directions
explicitly.

For propagation along the $z$-axis, we have $k^0=E$ and $k^3\equiv
k\simeq E(1-m^2/2E^2)$ and we choose $y=0,\, x=b
$. We find
\begin{eqnarray}\label{firstcontrib}
  -i\frac{k^\lambda}{m}\bar{\nu}_R\Gamma_\lambda\nu_L&=&\frac{k}{m}\varphi_{,1}+i\frac{m}{4E}h_{2,3}\,\,,\\
    \frac{k^\lambda k_\mu
    }{2mE}\, (h^\mu_{{\hat \alpha},\,
    \lambda}+\Gamma^\mu_{\alpha\lambda})\nu_R^\dagger \gamma^{\hat
    \alpha}\nu_L&=&-\frac{k}{2m}\left(1+\frac{k^2}{E^2}\right)\frac{GM}{2b}\,.\nonumber
\end{eqnarray}
Summing up, and neglecting terms of $ \mathcal{O}((m/E)^2)$,
Eq.(\ref{RTL}) becomes
\begin{equation}\label{RTL1}
\langle\nu_R|\dot{x}^{\mu}\partial_{\mu}\hat{T}|\nu_L\rangle=\frac{1}{2}\varphi_{,1}+\frac{i}{4}h_{2,3}\,.
\end{equation}
As a consequence
\begin{equation}\label{dadz}
  \frac{d\alpha}{dz}\simeq\frac{m}{E}\frac{d\alpha}{d\lambda}=
  \frac{m}{E}\left(\frac{1}{2}\varphi_{,1}+\frac{i}{4}h_{2,3}\right)\,,
\end{equation}
and the probability amplitude for the $ \nu_{L}\rightarrow\nu_{R}$
transition is of ${\cal O}(m/E)$, as expected.

Integrating (\ref{dadz}) from $-\infty$ to $z$, yields
\begin{eqnarray}\label{alfa-z}
  \alpha&\simeq&\frac{m}{E}\left[\frac{1}{2}\int_{-\infty}^zdz\varphi_{,1}+\frac{i}{4}h_2(z)\right]\\
  &=&\frac{m}{E}\frac{GM}{2b}\left[1+\frac{z}{r}-i\frac{2\omega
  R^2b^2}{5r^3}\right]\,.\nonumber
\end{eqnarray}
It also follows that
\begin{equation}\label{|alfa-z|}
  P_{L\rightarrow R}(-\infty,z)\simeq \left(\frac{m}{E}\right)^2\left(\frac{GM}{2b}\right)^2
  \left[\left(1+\frac{z}{r}\right)^2+\left(\frac{2\omega b^2R^2}{5r^3}\right)^2\right]\,.
\end{equation}
In this qubit, the first term in (\ref{|alfa-z|}) comes from the mass
of the gravitational source. The second from the source's angular
momentum and vanishes for $r\rightarrow\infty$ because  the
contribution from $-\infty $ to 0 exactly cancels that from 0 to $
+\infty $. In fact, if we consider neutrinos propagating from 0 to
$ +\infty $, we obtain
\begin{equation}\label{alfa infinito}
  P_{L\rightarrow R}(0,+\infty)\simeq \left(\frac{m}{E}\right)^2\left(\frac{GM}{2b}\right)^2
  \left[1+\left(\frac{2\omega R^2}{5b}\right)^2\right]\,.
\end{equation}
According to semiclassical spin precession equations
\cite{montague}, there should be no spin motion because spin and $
\vec{\omega}$ are parallel. The probabilities (\ref{|alfa-z|}) and
(\ref{alfa infinito}) mark therefore a departure from expected
results. They yield however results that are small of second order. Both expressions
vanish for $ m \rightarrow 0$, as it should because helicity is conserved \cite{mobed}.
It is interesting to observe that spin precession also occurs when
$ \omega$ vanishes \cite{aldov,casini}. In the case of
(\ref{|alfa-z|}) the mass contribution is larger when $ b<
(r/R)\sqrt{\frac{5r}{2\omega}}$, which, close to the source, with
$b\sim r\sim R$, becomes $ R\omega < 5/2$ and is always satisfied.
In the case (\ref{alfa infinito}), the rotational
contribution is larger if $ b/R < 2\omega R/5$ which restricts the region of dominance to a strip about the $
z$-axis in the equatorial plane, if the source is compact and $
\omega$ is relatively large.

In proximity of the source where the gravitational field is stronger
and $r \sim b\sim R$ the evolution equations for $\alpha$ and $\beta$
are $\frac{d\alpha}{d\lambda}\approx \tilde{D} \beta$ and
$\frac{d\beta}{d\lambda}\approx \tilde{D}^{*}\alpha$,
$\tilde{D}$ is almost constant and $\alpha$ and $\beta$ oscillate with
frequency
\begin{equation}\label{OM}
\Omega=\sqrt{\tilde{D}\tilde{D}^{*}}\approx \frac{m}{E}\frac{GM}{2b^2}\left\{1+\left(\frac{6\omega b}{5}\right)^2 \right\}^{\frac{1}{2}}\,.
\end{equation}
The contribution of the source rotation is therefore $(\omega b/c)^2\sim (v/c)^2$.
The helicity oscillations discussed are, in principle, relevant in astrophysics because
right-handed neutrinos are considered sterile. This point is discussed in the next section.

When propagation is along $x$, we put $k^0=E$, $k^1\equiv k\simeq E(1-m^2/2E^2)$.
The calculation can be simplified by assuming
that the motion is in the equatorial plane with $z=0$, $y=b$. We
then have
\begin{eqnarray}\label{contribx}
-i\frac{k^\lambda}{m}\bar{\nu}_R\Gamma_\lambda\nu_L&=&i\frac{k}{m}\varphi_{,2}+i\frac{E^2+k^2}{4mE}h_{1,2}-i\frac{E^2-k^2}{4mE}h_{2,1}\,\,,\\
    \frac{k^\lambda k_\mu
    }{2mE}\, (h^\mu_{{\hat \alpha},\,
    \lambda}+\Gamma^\mu_{\alpha\lambda})\nu_R^\dagger \gamma^{\hat
    \alpha}\nu_L&=&-i\frac{k}{2m}\left(1+\frac{k^2}{E^2}\right)\varphi_{,2}-i\frac{k^2}{2mE}h_{1,2}\,.\nonumber
\end{eqnarray}
Summing up, and neglecting terms of $ {\cal O}(m/E)^2$,
Eq.(\ref{RTL}) becomes
\begin{equation}\label{RTL1x}
\langle\nu_R|\dot{x}^{\mu}\partial_{\mu}\hat{T}|\nu_L\rangle=\frac{i}{2}\varphi_{,2}+\frac{i}{4}(h_{1,2}-h_{2,1})\,.
\end{equation}
The contributions to ${\cal O}((E/m)^2)$ again vanish and we get
\begin{equation}\label{dadx}
\frac{d\alpha}{dx}\simeq\frac{m}{E}\frac{d\alpha}{d\lambda}=\frac{m}{E}\left[\frac{i}{2}
\varphi_{,2}+\frac{i}{4}(h_{1,2}-h_{2,1})\right]\sim{\cal
O}(m/E)\,.
\end{equation}
Integrating (\ref{dadx}) from $-\infty$ to $x$, we obtain
\begin{equation}\label{alfax}
  \alpha\simeq i\frac{m}{E}\frac{GM}{2b}\left(1-\frac{2\omega
  R^2}{5b}\right)\left(1+\frac{x}{r}\right)\,
\end{equation}
and
\begin{equation}\label{alfa2x}
  P_{L\rightarrow
  R}(-\infty,x)\simeq\left(\frac{m}{E}\right)^2\left(\frac{GM}{2b}\right)^2\left(1-\frac{2\omega
  R^2}{5b}\right)^2\left(1+\frac{x}{r}\right)^2\,.
\end{equation}
Obviously, the contribution of $M$ is the same as for $z$-axis propagation.
However, the two cases differ substantially in
the behaviour of the term containing $\omega$. In this case, in fact,
the term does not vanish for $r\rightarrow\infty$.
If we consider neutrinos generated at $x=0$ and propagating to
$x=+\infty$, we find
\begin{equation}\label{pinfinitox}
P_{L\rightarrow
R}(0,+\infty)\simeq\left(\frac{m}{E}\right)^2\left(\frac{GM}{2b}\right)^2\left(1-\frac{2\omega
  R^2}{5b}\right)^2\,.
\end{equation}
The $M$ term is larger when $ \frac{2\omega R^{2}}{5b}< 1$. At
the poles $ b\sim R$ and the $M$ term dominates because the
condition $ \omega R <5/2$ is always satisfied. The angular
momentum contribution prevails in proximity of the equatorial
plane. The transition probability vanishes at $ b=2\omega
R^{2}/5$.

An altogether different type of qubit is represented by neutrino flavour oscillations.
They have been discussed in the context of the Lense-Thirring metric in \cite{PAP5}.
The qubit frequency is in this case
proportional to $\Delta m^{2} =m_{2}^{2}-m_{1}^{2}$, where $m_{1}$ and $m_{2}$
are the masses of the neutrino mass eigenstates.

\section{5. Neutrino conversion in supernovae}

\hspace*{5mm}
The results of the previous section may be applied to the propagation of a beam of
neutrinos {\it in vacuo}. The presence of a medium is realized by means of a potential $V$.
The neutrinos are massive and may
therefore have a magnetic moment $\mu$. In the presence of an external
magnetic field $\vec{B}$ and of the Mashhoon term proportional to the angular
velocity of the source $\vec{\Omega}$, the evolution equations
become \cite{papprd1,papini94}
\begin{eqnarray}
i  \frac{d \nu}{dz} & = & \frac{m^{2} }{2E} - \frac{1}{2}
\left( P_{+} \vec{\Omega} \cdot \vec{\sigma}  P_{-} +
P_{-} \vec{\Omega} \cdot \vec{\sigma}  P_{+} \right) \nu \nonumber \\
&  & - \mu\left( P_{+} \vec{B} \cdot \vec{\sigma}  P_{-} +
P_{-} \vec{B} \cdot \vec{\sigma} P_{+}\right) \nu,
\label{eq5.1}
\end{eqnarray}
where $\nu = \left( \begin{array}{c} \nu_{R} \\ \nu_{L} \end{array} \right)$
and $P_{\pm}=(1\pm \sigma_{3})/2$ are the $R$ and $L$ projection operators.
Eq.(\ref{eq5.1}) leads to neutrino oscillations.

The frequency of oscillation is then
$\Omega_{\perp}/2$ where $\Omega_{\perp}$ is the component of the
angular velocity normal to the neutrino trajectory. In particular, if a beam
of neutrinos consists of $N_{L}(0)$ particles at $z=0$, the relative numbers
of $\nu_{_{L}}$ and $\nu_{_{R}}$ at $z$ will be
\begin{eqnarray}
N_{L}(z) = N_{L}(0) \cos^{2}\left(\frac{ \Omega_{\perp}z}{2} \right),
\hspace*{5mm}
N_{R}(z) = N_{R}(0) \sin^{2}\left(\frac{\Omega_{\perp}z}{2} \right),
\label{eq5.2}
\end{eqnarray}
These oscillations are interesting because the $\nu_{_{R}}$'s,
if they exist, do not interact. They would therefore provide an energy
dissipation mechanism with possible astrophysical implications. The
conversion rate is not large for galaxies and white dwarfs. In fact one
can obtain from (\ref{eq5.2}) $N_{_{R}} \sim 10^{-6}
N_{_{L}}(0)$ for galaxies of size $L$ for which $\Omega_{\perp}L
\sim 200\,$km/s. Similarly, for white dwarfs for which $\Omega_{\perp}
\sim 1.0\,s^{-1}$, one finds $N_{_{R}} \sim \,10^{-4} N_{_{L}}(0)$. On the other
hand, the $\nu_{_{L}}$'s diffuse out of a canonical neutron star in a time
$1$ to $10\,$s, during which they travel a maximum distance $3\times 10^{9}\,$cm
between collisions. This and the fact that for a millisecond pulsar
the conversion rate $\nu_{_{L}} \rightarrow \nu_{_{R}}$ is $\sim 0.5$ at
distances $L \sim 5 \times 10^{6}\,$cm suggest that the dynamics of the star
could be affected by such a cooling mechanism. Indeed the star may even cool
too rapidly at higher rotational speeds for a pulsar to form.

The magnetic moment of the neutrino does not appear in
the calculations because magnetic spin-flip rates of
magnitude comparable to (\ref{eq5.2}) would
require magnetic moments in excess of the value $\mu \sim
10^{-19} \mu_{_{B}} \left(\frac{m_{\nu}}{1eV} \right)$ predicted by the
standard model.

\hspace*{5mm}
The behaviour of neutrinos in a medium is modified by a potential $V$ that
vanishes for $\nu_{_{R}}$'s. In the core of a supernova $V$ can be written
as \cite{Kainulainen1}
\begin{eqnarray}
V(\nu_{e}) = 14 \mbox{eV} \frac{\rho}{\rho_{c}} y(\vec{r}, t),
\label{eq6.1}
\end{eqnarray}
where $y(\vec{r}, t) \equiv 3 Y_{e}(\vec{r}, t) + 4 Y_{\nu_{e}}(\vec{r}, t)
- 1$, the $Y$'s represent the lepton fractions present and $\rho_{c} =
4 \times 10^{14}$g/cm$^{3}$. For supernovae $V$ can be large,
of the order of several electron volts, and rotation may be neglected. Only the acceleration term in (\ref{eq5.2}) need be
considered and the effective Hamiltonian then has the form \cite{papini94}
\begin{eqnarray}
H = \left|
\begin{array}{cc}
V   &   {\displaystyle \frac{\hbar a_{\perp}}{2c}}  \\
{\displaystyle \frac{\hbar a_{\perp}}{2c}}  &
{\displaystyle \frac{c^{4}\delta m^{2}}{2E}}
\end{array}
\right|,
\label{eq6.2}
\end{eqnarray}
where $\delta m^{2} \equiv
m_{\nu_{_{L}}}^{2} - m_{\nu_{_{R}}}^{2}$ and $a_{\perp}$ is the component
of the acceleration transverse to the neutrino trajectory. If the initial
state is pure $\nu_{_{L}}$ and the number of particles in this state is
$N_{0}$ at $z = 0$, then the corresponding numbers of $\nu_{L}$ and $\nu_{R}$ at $z$ are
\begin{eqnarray}
N_{L} = N_{0} \left[ \cos^{2} (\tilde{\Omega} z) + \cos^{2}(2 \theta_{a})
\sin^{2} (\tilde{\Omega} z) \right],
\hspace*{5mm}
N_{R} = N_{0} \sin^{2} (2 \theta_{a}) \sin^{2} (\tilde{\Omega} z),
\label{eq6.3}
\end{eqnarray}
where
\begin{eqnarray}
\sin^{2}(2 \theta_{a}) \equiv \frac{
\left( {\displaystyle \frac{\hbar a_{\perp}}{2c}} \right) ^{2}
}{
\left( {\displaystyle \frac{\hbar a_{\perp}}{2c}} \right) ^{2} +
4 \left( V - {\displaystyle \frac{c^{4}
\delta m^{2}}{2E}} \right)
};
\hspace*{3mm}
\tilde{\Omega} \equiv \sqrt{
\frac{1}{c^{2}} \left( \frac{a_{\perp}}{2c} \right) ^{2} +
\frac{1}{4 \hbar^{2} c^{2}} \left( V - \frac{c^{4}
\delta m^{2}}{2E} \right) ^{2} }.
\label{eq6.4}
\end{eqnarray}
If $\frac{1}{2} \left( V - \frac{c^{4} \delta m^{2}}{2E} \right) >
\frac{\hbar a_{\perp}}{2c}$, spin precession is strongly suppressed and the
flux of particles at $z$ consists mainly of $\nu_{_{L}}$'s. The conversion
takes place at resonance if $V = \frac{c^{4} \delta m^{2}}{2E}$.

Summarizing, the components of acceleration transverse to the particle path
couple to its spin. This and the Mashhoon term
applied to massive neutrinos, produce $\nu_{_{L}} \leftrightarrow
\nu_{_{R}}$ oscillations which may have macroscopic effects if the
$\nu_{_{R}}$'s are sterile, as frequently assumed. In fact
$\nu_{_{L}} \rightarrow \nu_{_{R}}$ conversion by rotation-spin
coupling may help to explain why pulsars of period shorter than a
millisecond are relatively rare.

\section{6. Spin currents}

 The realization that the flow of spin angular momentum can be separated from that of charge has
recently stimulated intense interest in fundamental spin physics \cite{ziese}, particularly in view of its
applications \cite{igor,bauer,KYM}.

In this section we study the generation and control of spin currents by rotation and acceleration \cite{PASP}.
In this context the fundamental tool still is the covariant Dirac equation.

We use the first order solutions of (\ref{DiracEquation}) that have the form
\begin{equation}\label{E}
  \Psi(x) = {\hat T}(x) \Psi_{0}(x)\,,
\end{equation}
where $\Psi_{0}(x)$ is a solution of (\ref{DE}) and the operator $\hat{T}$ is given by (\ref{PsiSolution}), or (\ref{PsiSolution2}).

When acceleration and rotation are present, $\gamma_{\mu\nu}$ is given by \cite{hehl.ni,singh}
\begin{equation} \label{met}
\gamma_{00}\approx 2({\bf a}\cdot {\bf x})+({\bf a}\cdot {\bf x})^2-{\bf \Omega}^{2} {\bf x}^{2} +({\bf \Omega}\cdot {\bf x})^{2}\,, \\
\gamma_{0i}=-({\bf \Omega}\times {\bf x})_i\,,\gamma_{ij}=\eta_{ij}\,,
\end{equation}
where ${\bf a} $ and $ {\bf\Omega} $ represent acceleration and rotation respectively. To first order the tetrad is given by
\begin{eqnarray} \label{tet}
e^{\mu}_{\hat{\alpha}}&\approx &\delta^{\mu}_{\alpha}+h^{\mu}_{\hat{\alpha}}\,, h^{0}_{\hat{0}}=-{\bf a}\cdot {\bf x}\,,
h^{0}_{\hat{i}}=0\,, h^{k}_{\hat{i}}=0\,, h^{i}_{\hat{0}}=-\varepsilon^{ijk}\Omega_j x_k\,,\\
h^{\hat{0}}_{0}&=& {\bf a}\cdot {\bf x}\,, h^{\hat{k}}_{0}=\varepsilon_{ijk}\Omega^{i}x^{j}\,, h^{\hat{0}}_{i}=0, h^{\hat{k}}_{i}=\delta^{k}_{i} \nonumber \,.
\end{eqnarray}
from which the spinorial connection can be calculated in the usual way.
The result is $\Gamma_i =0$ and $\Gamma_0 =-\frac{1}{2}a_{i}\sigma^{\hat{0}\hat{i}}-\frac{1}{2}{\bf \Omega}\cdot{\bf \sigma}I$.

For electrons, $u_1$ corresponds to the choice $\phi =\phi_{1}$ and $u_2$ to $\phi =\phi_2$. Substituting in (\ref{psi0}), one finds the
spinors $u_{1}$ and $u_{2}$.
These are not eigenspinors of the matrix $\Sigma^3 =\sigma^3 I$ and do not, therefore, represent the spin components in
the $z$-direction. They become however eigenspinors of $\Sigma^3$ when $k^1 =k^2 =0$, or when $\textbf{k}=0$ (electron rest frame).

The appropriate way to determine whether there is transfer of angular momentum between the external non-inertial field and the electron spin is to use the
third rank spin current tensor \cite{grandy}
\begin{equation}\label{ST}
S^{\rho\mu\nu}=\frac{1}{4im}\left[\left(\nabla^{\rho}\bar{\Psi}\right)\sigma^{\mu\nu}(x)\Psi-\bar{\Psi}\sigma^{\mu\nu}(x)\left(\nabla^{\rho}\Psi\right)\right]\,,
\end{equation}
that in Minkowski space satisfies the conservation law $S^{\rho\mu\nu},_{\rho}=0 $ when all $\gamma_{\alpha\beta}(x)$ vanish and yields in addition the expected
result $S^{\rho\mu\nu}= S^{0\mu\nu}$ in the rest frame of the particle. Writing $\sigma^{\mu\nu}(x)\approx \sigma^{\hat{\mu}\hat{\nu}}+h^{\mu}_{\hat{\tau}}\sigma^{\hat{\tau}\hat{\nu}}+h^{\nu}_{\hat{\tau}}\sigma^{\hat{\mu}\hat{\tau}}$, using the relation $ \Phi_{G,\mu\nu}=k_{\alpha}\Gamma^{\alpha}_{\mu\nu}$ and substituting (\ref{E}) and (\ref{PsiSolution2}) in (\ref{ST})
one obtains, to $\mathcal{O}(\gamma_{\alpha\beta})$,
\begin{equation}\label{STF}
S^{\rho\mu\nu}=\frac{1}{16im^3}\bar{u}_{0}\left\{8im^2k^{\rho}\sigma^{\hat{\mu}\hat{\nu}}+8imk^{\rho}h^{[\mu}_{\hat{\tau}}\sigma^{\hat{\tau}\hat{\nu}]}+ \right.
\end{equation}
\[
\left.
4imk^{\rho}\left(\Phi_{G,\alpha}
+k_{\sigma}h^{\sigma}_{\hat{\alpha}}\right)\left\{\sigma^{\hat{\mu}\hat{\nu}},\gamma^{\hat{\alpha}}\right\}-8imk^{\rho}\Phi_{G}k^{[\mu}\gamma^{\hat{\nu}]}+ \right.
\]
\[
\left.
4mk^{\rho}k_{\alpha}\left[\sigma^{\hat{\mu}\hat{\nu}},\left(\gamma^{\hat{\alpha}}\Phi_{S}-\gamma^{\hat{0}}\Phi^{+}_{S}
\gamma^{\hat{0}}\gamma^{\hat{\alpha}}\right)\right]+
4m^2k^{\rho}\left[\sigma^{\hat{\mu}\hat{\nu}},\left(\Phi_{S}-\gamma^{\hat{0}}\Phi^{+}_{S}\gamma^{\hat{0}}\right)\right]- \right.
\]
\[
\left. 8m^2 k^{\rho}h^{0}_{\hat{\alpha}}\left[\gamma^{\hat{0}},\left[\sigma^{\hat{0}\hat{\alpha}},\sigma^{\hat{\mu}\hat{\nu}}\right]\right] -8im^2k_{\sigma}\left(\Gamma^{\sigma}_{\alpha\beta}\eta^{\beta\rho}+\partial^{\rho}h^{\sigma}_{\hat{\alpha}}\right)\eta^{\alpha[\mu}\gamma^{\hat{\nu}]}+ \right.
\]
\[
\left. 8im^2\partial^{\rho}\Phi_{G}\left(4m\sigma^{\hat{\mu}\hat{\nu}}-2ik^{[\mu}\gamma^{\hat{\nu}]}\right)+4im^2\gamma^{\hat{0}}\Gamma^{\rho+}\gamma^{\hat{0}}\left\{\left(\gamma^{\hat{\alpha}}
k_{\alpha}+m\right),\sigma^{\hat{\mu}\hat{\nu}}\right\}\Gamma^{\rho} \right\}u_{0}
\,. \]
It is therefore possible to separate $S^{\rho\mu\nu}$ in inertial and non-inertial parts. The first term on the r.h.s. of (\ref{STF}) gives the result expected when $\vec{k}=0$ and the external field vanishes. From (\ref{STF}) one finds
\begin{equation}\label{CST}
\partial_{\rho}S^{\rho\mu\nu}=\frac{1}{16im^3}\bar{u}_{0}\left\{8imk^{\rho}\partial_{\rho}h^{[\mu}_{\hat{\tau}}\sigma^{\hat{\tau}\hat{\nu}]}
-8imk^{\rho}\Phi_{G,\rho}k^{[\mu}\gamma^{\hat{\nu}]}+
\right.
\end{equation}
\[
\left.
4imk^{\rho}\left(k_{\sigma}\Gamma^{\sigma}_{\alpha\rho}+\partial_{\rho}h^{\sigma}_{\hat{\alpha}}k_{\sigma}\right)
\left\{\sigma^{\hat{\mu}\hat{\nu}},\gamma^{\hat{\alpha}}\right\}
+4mk^{\rho}k_{\alpha}\left[\sigma^{\hat{\mu}\hat{\nu}},\left(\gamma^{\hat{\alpha}}\Gamma_{\rho}-\gamma^{\hat{0}}\Gamma^{+}_{\rho}
\gamma^{\hat{0}}\gamma^{\hat{\alpha}}\right)\right]+ \right.
\]
\[
\left. 4m^2k^{\rho}\left[\sigma^{\hat{\mu}\hat{\nu}},\left(\Gamma_{\rho}-\gamma^{\hat{0}}\Gamma^{+}_{\rho}\gamma^{\hat{0}}\right)\right]+
8m^2 k^{\rho}\partial_{\rho}h^{0}_{\hat{\alpha}}\left[\gamma^{\hat{0}},\left[\sigma^{\hat{0}\hat{\alpha}},\sigma^{\hat{\mu}\hat{\nu}}\right]\right]- \right.
\]
\[
\left. 8im^2k_{\sigma}\partial^{\rho}\Gamma^{\sigma}_{\alpha\rho}\eta^{\alpha[\mu}\gamma^{\hat{\nu}]}+8im^2k_{\sigma}\Gamma^{\sigma}_{\rho\tau}\eta^{\tau\rho}
\left(4m\sigma^{\hat{\mu}\hat{\nu}}-2ik^{[\mu}\gamma^{\hat{\nu}]}\right)+ \right.
\]
\[
\left. 8im^2k^{\alpha}\Gamma^{\rho}_{\alpha\rho}\sigma^{\hat{\mu}\hat{\nu}}+8im^2k^{\rho}\Gamma^{\mu}_{\alpha\rho}\sigma^{\hat{\alpha}\hat{\nu}}-8im^2k^{\rho}\Gamma^{\nu}_{\alpha\rho}\sigma^{\hat{\alpha}\hat{\mu}}\right\}u_{0}\,, \]
where terms containing $\Gamma_{0,0}=0$ and $\partial_{\alpha}\partial_{\beta}h^{\mu}_{\hat{\nu}}=0$ have been eliminated. It therefore follows that the external field invalidates the simple conservation law $\partial_{\rho}S^{\rho\mu\nu}=0$ and that there is in this case continual interchange between spin and orbital angular momentum. The result is entirely similar to that found for external electromagnetic fields \cite{grandy}.
Thus, in principle, one can use non-inertial fields to generate spin currents. In the rest frame
of the particle and when $ {\bf \Omega}=(0,0,\Omega)$, one finds
\begin{equation}\label{RF}
\partial_{\rho}S^{\rho\mu\nu}=\partial_{i}S^{i12}=\frac{1}{2}\left(\Gamma^{\rho}_{0\rho}+\Gamma^{1}_{10}+\Gamma^{2}_{20}\right)\bar{u}_{0}\sigma^{\hat{1}\hat{2}}u_{0}=\frac{E+m}{2E}\frac{\Omega a_{2} x}{1+{\bf a}\cdot {\bf x}}\,,
\end{equation}
and $ \partial_{i}S^{i13}= \partial_{i}S^{i23}=0$. In (\ref{RF}) $u_{0}$ corresponds to $u_{1}$. The direct coupling of the non-inertial field to the particle's spin current violates the law
$\partial_{\rho}S^{\rho\mu\nu}=0$. Conservation is restored if the parameters $\Omega$, or $a_{2}$,
or both vanish.

Qubits appear in the actual spin motion. In general, transfer of angular momentum between external and non-inertial fields occurs when the operator $\hat{T}$
has some non-diagonal matrix elements. If in fact at time $t=0$ a beam of electrons is entirely of the $u_{2}$ variety, at time $t$ the fraction of
$u_1$ is $|\langle u_1|\hat{T}|u_2\rangle|^2$. The last expression becomes, along the electron world line,
\begin{equation}\label{me}
  P_{2\rightarrow 1}=\left|\langle u_1|\hat{T}|u_2\rangle\right|^{2}
  =\left|\int_{\lambda_0}^\lambda\langle u_1| {\dot x}^\mu \partial_{\mu}{\hat
  T}|u_2\rangle d\lambda\right|^2\\,
\end{equation}
 where, as usual, $\dot{x}^{\mu}=k^{\mu}/m$ and $\lambda$ is the affine parameter along the world line. From \cite{PAP5}
 \begin{equation}\label{tnu}
 \partial_{\nu}\hat{T} =\frac{1}{2m}\left\{h^{\mu}_{\hat{\alpha,\nu}}\gamma^{\hat{\alpha}}k_{\mu}+ \gamma^{\hat{\mu}}\Phi_{G,\mu\nu}-2im \left(\Phi_{G,\nu}+\Gamma_{\nu}-e A_{\nu}\right)\right\}\,,
  \end{equation}
  one can see that
 \begin{equation}\label{tnu}
 \langle u_1|\frac{k^{\nu}}{m}\partial_{\nu}\hat{T}|u_2\rangle =
 -i\frac{k^{0}}{m} \langle u_1|\Gamma_{0}|u_2\rangle=
 -i\frac{k^{0}}{m}\langle u_1|\left\{-\frac{1}{2} a_{i}\sigma^{\hat{0}\hat{i}}-
 \frac{1}{2}\Omega_{i} \sigma^{i} I \right\} |u_2 \rangle .
 \end{equation}
 A useful way to visualise the spin motion under the action of rotation and acceleration follows from the Mashhoon term $H_{M}=-{\bf \Omega}\cdot {\bf s}$, where ${\bf s}=\frac{{\bf \sigma}}{2}$, and from $\frac{1}{2}a_{i}\sigma^{\hat{0}\hat{i}}=\frac{i}{2}a_{i}\alpha^{\hat{i}}$. The two interaction terms lead to the first order equation of motion \cite{jackson}
 \begin{equation}\label{spin}
 \frac{d{\bf s}}{dt}={\bf s}\times \left({\bf \Omega}+ {\bf v}\times{\bf a}\right)\,.
 \end{equation}
Note that $A_{\mu}$, introduced by
writing $\Phi_{T}=\Phi_{S}+\Phi_{G}+\Phi_{EM}$, where $\Phi_{EM}=e\int_{P}^{x}dz^{\lambda}A_{\lambda}(z)$, does not contribute to (\ref{tnu}) because $\langle u_1|u_2\rangle =0$. Note also that the terms $ie (h^{\mu}_{\hat{\alpha}}\gamma^{\hat{\alpha}}k_{\mu}+\gamma^{\hat{\mu}}\Phi_{G,\mu})A_{\nu}\frac{k^{\nu}}{m}$ drop out, in the particle rest frame, on account of $\langle u_1|\gamma^{\hat{\mu}}|u_2\rangle =0$.
No mixed effects of first order in rotation or acceleration and first order in the electromagnetic field are therefore present in this calculation. This applies to all terms containing the magnetic field ${\bf B}$, like the Zeeman term, and electric fields, like the spin-orbit interaction, that are present in the lowest order Dirac Hamiltonian that can be derived from (\ref{DiracEquation}) \cite{singh}. To $\mathcal{O}(\gamma_{\mu\nu})$, contributions to (\ref{tnu}) from the electromagnetic field are present in the actual determination of the electron's path, as stated above.

From (\ref{tnu}) one obtains
\begin{eqnarray}\label{res}
 \frac{2m}{iE}\langle u_1|\frac{k^{\nu}}{m}\partial_{\nu}\hat{T}|u_2\rangle &=&-i\frac{k^3}{E}a_1-\frac{k^3}{E}a_2 +i\frac{k^1 -ik^2}{E}a_3\\ &+& \Omega^3 \frac{k^3}{E}\frac{-k^1 +ik^2}{E+m}\nonumber \\ &+&
 \Omega^1 \frac{E+m}{2E}\left(1+\frac{(k^{3})^2}{(E+m)^2}-\frac{(k^1 -ik^2)^2}{(E+m)^2}\right)\nonumber \\ &-&
 i\Omega^2 \frac{E+m}{2E}\left(1+\frac{(k^{3})^2}{(E+m)^2}+\frac{(k^1 -ik^2)^2}{(E+m)^2}\right)\equiv A_{12} \nonumber  \,,
 \end{eqnarray}
where $k^0\equiv E$. The parameter $k_{\mu}$ corresponds to the electron four-momentum when $ {\bf \Omega}=0$ and $ {\bf a}=0$.

Some general conclusions can be drawn from (\ref{res}).

i) If $k^3 =0$, the particles move in the $(x,y)$-plane. If, in addition, $\Omega^1 =\Omega^2 =0$, then $A_{12}\neq 0$ only if $a_3 \neq 0$.

ii) If, however, ${\bf a}$ is also due to rotation, the conditions $\Omega_1 =\Omega_2 =0$ imply $a_3 =0$ and therefore $A_{12}=0$. This is the relevant case of motion in the $(x,y)$-plane with rotation along an axis perpendicular to it. One cannot therefore have a rotation induced spin current in this instance.

iii) Even for ${\bf k}=0$ one can have $A_{12}\neq 0$ if one of $\Omega_1$ and $\Omega_2$ does not vanish. This is a direct consequence of the spin rotation interaction or Mashhoon term contained in (\ref{DiracEquation}) \cite{MASH1,MASH2,singh}.

A few examples are discussed below.

Consider an electron wave packet moving along the $x$-axis of a frame rotating about the same axis. Then ${\bf a}=0$ and the remaining parameters are
${\bf k}= (k,0,0)$ and ${\bf \Omega}=(\Omega,0,0)$. While the electron propagates along $x$, $u_1$ and $u_2$ propagate in opposite directions along
$x$ because of (\ref{tnu}) and (\ref{spin}). For a beam the spin current generated by rotation is therefore $I_s =I_{\downarrow}-I_{\uparrow}$, with obvious
meaning of the symbols. One finds $P_{2\rightarrow 1}=|\frac{i\Omega(E+m)}{4m}\int_{0}^{t}dt\left(1-\frac{k^2}{(E+m)^2}\right)|^2=(\frac{\Omega t}{2})^2$ which holds for $t\leq 2/\Omega$
on account of the requirement $P_{2\rightarrow 1}\leq 1$.

 Consider next a wave packet moving in  the plane $z=0$ itself
  moving along $z$ with velocity $k^3 /E$, while rotating about the $z$-axis with ${\bf \Omega}=(0,0,\Omega)$. The other parameters are ${\bf k}=(k\cos\Omega t, k\sin \Omega t,k^3)$ and ${\bf a}=(-\Omega^2 x,-\Omega^2 y,0)$. From (\ref{res}) one finds $A_{12}=(\frac{k^3}{E})(-ia^1-a^2 +\Omega \frac{-k^1 +ik^2}{E+m})$ and $P_{2\rightarrow 1}= \left[\frac{k^3}{E}\left(\Omega R+\frac{k}{E+m}\right)\sin2\Omega t\right]^2$, where $R$ is the radius of the circle described by the wave packet in the plane $z=0$. The motion of the center of mass of the wave packet is helical and so is the motion of the spin components which propagate, however, in opposite directions giving rise to a spin current.

In the case of motion occurring in the plane $z =0$ ($k^3=0$) and
${\bf \Omega}=(\Omega,0,0)$, one also gets ${\bf a} = (-\Omega^2 x,0,0)$ and
\begin{equation}\label{rot}
A_{12}=  \Omega \frac{E+m}{2E}\left\{1-\frac{(k^1-ik^2)^2}{(E+m)^2}\right\}\,,
\end{equation}
where $k^1 =k \cos\omega t$, $k^2 = k \sin \omega t$ and $\omega = \frac{eB}{m\gamma}$ is the cyclotron frequency of the electrons along the circular path determined by the constant magnetic field $B$. Substituting (\ref{rot}) in (\ref{me}) one finds
\begin{eqnarray}\label{rot1}
P_{2\rightarrow 1}&=&\left|i \Omega \frac{E+m}{2E}\int^{t}_{0}\left\{1-\frac{k^2 (\cos 2\omega t -i\sin 2\omega t)}{(E+m)^2}\right\}dt\right|^2 \\&=&
\left(\Omega t \frac{E+m}{4E}\right)^2 \left\{\frac{\sin^4 \omega t}{\omega^2 t^2}+\left(1-\frac{k^2}{(E+m)^2}\frac{\sin 2\omega t}{2\omega t}\right)^2\right\}\nonumber \,,
\end{eqnarray}
which holds for all $ t $ for which $P_{2 \rightarrow 1}\leq 1 $.
While both spin up and down electrons move on a circle of radius $R$ about the $z$-axis, they propagate in opposite directions because of the spin-rotation coupling, thus generating a spin current.

Summarizing, rotation and acceleration can be used to generate and control spin currents. This follows from the covariant Dirac equation and its
exact solutions to $\mathcal{O}(\gamma_{\mu\nu})$.
To this order, external electromagnetic fields can be taken into account through the particle motion.
The transition amplitude for the conversion spin-up to spin-down is proportional to $A_{12}$ and is expressed as a
function of ${\bf \Omega}$, ${\bf a}$ and of the electron four-momentum $k_{\mu}$ before the onset of rotation and acceleration.
The same expression suggests criteria for the generation of spin currents and the transfer of momentum and
angular momentum to them. No energy can, of course, be transferred from the non-inertial fields to the spin currents as long as $\gamma_{\mu\nu}$ remains stationary.

The particular forms of (\ref{res}) discussed above provide additional examples of gravitational qubits.

\section{7. Summary}

Gravitational qubits are the simplest quantum systems that can be employed
to study gravity. They exist in the laboratory and astrophysical conditions.
Attention has been focussed on spin-$1/2$ fermions because of their simple
eigenstate structure. Spin-flip transitions occur in nature frequently.
We have considered here some of those that are characterized by sustained oscillations.

To the laboratory belong particles rotating in accelerators.
Their quantum beats refer to spin oscillations mediated by the Mashhoon
spin-rotation interaction. They have been observed and play an important role
in important measurement of the anomalous magnetic moment of the muon.

Section 3 is entirely devoted to the calculation of the deflection of
fermions in a gravitational background described by the Lense-Thirring metric.
The interest is limited here to the action of rotation on the particle spin.
The procedure can be also applied to bosons. Basically, the part of the deflection
that does not depend on rotation is that predicted by general relativity. Rotation of the source
yields in general additional corrections which are due to the particle spin and are
therefore quantum mechanical. These, for instance,
are present in the deflection equations (\ref{delta0z}) and (\ref{delta0x}), but with a
noticeable difference: in the case of (\ref{delta0x}) the contribution of the source
angular momentum does not vanish in the limit $x\rightarrow \infty$.

The neutrino helicity oscillations in vacuo described in Section 4 are small, but
intriguing because $\nu_{L}$ evolve into $\nu_{R}$ which are sterile.
This energy dissipation mode could be relevant to compact astrophysical objects.
The introduction in Section 5 of a medium in which neutrinos propagate
increases the $\nu_{_{L}} \rightarrow \nu_{_{R}}$ transition probability if the medium potential is attuned to
the difference of the mass squared of the neutrino mass eigenstates. The fact that
pulsars of period shorter than a millisecond are rare, lends support to the dissipation
mechanism implied by the $\nu_{_{L}} \rightarrow \nu_{_{R}}$ conversion by rotation-spin
coupling.

Spin currents and spin motion are introduced in Section 6. It is shown that
gravitational fields violate the usual formulation of the spin conservation law.
Some particular qubits are discussed together with the associated spin currents.
We also consider the case of fermions accelerated to the maximal acceleration limits contemplated
by Caianiello \cite{CAI1,CAI2,CAI3}. The calculations \cite{MAPAP}
confirm that continual interchange between spin and angular momentum can occur in this instance,
but only if the acceleration is time-dependent. This requires a transfer of energy from
a very compact star, or a black hole and the particle.
Even in the case, uniform acceleration produces no observable
effects on the particle spin, in agreement with \cite{BINI}.

Near-neighbour momenta oscillations of a gas of particles in
a gravitational field are discussed in \cite{magnons}.

\end{document}